# Lasing in Si$_3$N$_4$-organic hybrid (SiNOH) waveguides

DARIA KOHLER,[1,2,3] ISABEL ALLEGRO,[2] SENTAYEHU FETENE WONDIMU,[1,2] LOTHAR HAHN,[2] WOLFGANG FREUDE,[1] AND CHRISTIAN KOOS[1,2,4]

[1]*Institute of Photonics and Quantum Electronics (IPQ), Karlsruhe Institute of Technology (KIT), Engesserstrasse 5, 76131 Karlsruhe, Germany*
[2]*Institute of Microstructure Technology (IMT), Karlsruhe Institute of Technology (KIT), Hermann-von-Helmholtz-Platz 1, 76344 Eggenstein-Leopoldshafen, Germany*
[3]*daria.kohler@kit.edu*
[4]*christian.koos@kit.edu*

**Abstract:**
Lasers that operate in the visible wavelength range and can be easily integrated on a chip together with photonic integrated circuits are of particular interest for biophotonic applications. We present a new class of laser sources that emit at 600 nm wavelength and can be efficiently integrated on the Si$_3$N$_4$ platform using low-cost fabrication techniques. The Si$_3$N$_4$-organic hybrid lasers are optically pumped from top without the need of a precise alignment. We investigate different laser geometries based on spiral shaped ring resonators and distributed feedback resonators. The lasers show threshold fluences of $(40...70)\,\mu J/cm^2$. We discuss the laser properties with a focus on their application in the field of bioscience.

## 1. Introduction

Integrated sensors based on optical waveguides (WG) have an enormous application potential in biophotonics and in biomedical diagnostics, especially when it comes to multiplexed, highly sensitive detection of a wide variety of target molecules [1,2]. For biophotonic applications, operation in the visible wavelength range is of particular interest. The low absorption in aqueous solutions allows long interaction lengths in refractive-index sensors [3], leading to low-background absorption spectroscopy with large dynamic range [4], or to living-cell imaging using waveguide-based total internal reflection fluorescence (TIRF) microscopy [5]. Specifically, absorption in typical aqueous analytes, such as urine or saliva, is approximately three orders of magnitude smaller at visible (VIS) wavelength of, e. g., 600 nm than at near-infrared (NIR) wavelengths of, e. g., 1550 nm [6]. Similarly, blood analysis in photonic sensors largely relies on the so-called therapeutic window between 600 nm and 1100 nm, which offers a good compromise between pronounced haemoglobin absorption at shorter wavelengths and strong water absorption at higher wavelengths [7].

In this context, silicon nitride (Si$_3$N$_4$) has emerged as a highly attractive integration platform for waveguide-based sensors, offering low-loss propagation over a wide spectral range from VIS to NIR wavelengths [8-10]. Silicon nitride WG feature a large refractive-index contrast between the core ($n_{Si_3N_4} = 2.04$ @ $\lambda = 600$ nm) and the silicon dioxide cladding ($n_{SiO_2} = 1.64$ @ $\lambda = 600$ nm) or the aqueous analyte ($n_{H_2O} = 1.33$ @ $\lambda = 600$ nm), and are therefore perfectly suited for densely integrated sensor arrays with small footprint and low analyte consumption. Moreover, Si$_3$N$_4$-based photonic integrated circuits (PIC) can be efficiently fabricated in large quantities using mature wafer-scale processes that exploit the CMOS technology base, and for which a world-wide ecosystem of photonic foundries exists [8,9]. This opens a path towards cost-efficient mass production of highly functional sensor chips for one-time use in point-of-care diagnostics.

However, despite its various advantages, the basic Si$_3$N$_4$ integration platform is so far limited to passive building blocks. Silicon nitride PIC hence have to rely either on external light sources that are coupled to the chip [3,11,12], or on a hybrid integration of light-emitting materials in combination with passive $Si_3N_4$ WG [13-15]. Light supplied by external sources requires delicate fiber-chip coupling schemes that are subject to stringent mechanical tolerances, and the associated complexity is prohibitive for low-cost disposable sensors. Moreover, grating couplers, which are the mainstay for fiber-chip coupling in silicon photonics, suffer from low refractive-index contrast when realized on the $Si_3N_4$ platform. This leads to low grating strength, poor directionality, and low coupling efficiencies [16]. These challenges can be overcome by on-chip light sources. Previous demonstrations of visible-light sources on the $Si_3N_4$ platform comprise ring and disk resonators that are vertically coupled to $Si_3N_4$ WG cores and contain light-emitting perovskite materials [17,18] or CdS/CdSe quantum dots [19]. However, these approaches require complex manufacturing processes, such as material deposition from the gas phase and subsequent lithographic structuring. These techniques are not compatible with cost-efficient mass production through standard CMOS-based process workflows, which would be mandatory for disposable point-of-care sensors.

In this paper we demonstrate a new class of laser sources that can be efficiently integrated on the $Si_3N_4$ platform using low-cost fabrication techniques and do not require any lithographic structuring. The devices combine in a hybrid way nanophotonic $Si_3N_4$ WG cores and dye-doped organic cladding materials, which can be optically pumped by an external laser or a light-emitting diode (LED) without any high-precision alignment of the pump spot, and without any mechanical contact to the chip. The dye-doped organic cladding can be efficiently deposited on the wafer by spin coating, by local dispensing, or by inkjet printing. An appropriate choice of laser dyes allows the co-integration of $Si_3N_4$-organic hybrid (SiNOH) lasers with various emission wavelengths across the VIS and NIR spectral ranges. In our work, we develop a theoretical model of SiNOH lasers for spiral-shaped ring resonators and for distributed-feedback (DFB) resonators and we experimentally demonstrate lasing in both structure types. To the best of our knowledge, this is the first demonstration of lasers on the $Si_3N_4$ platform that exploits light emission in active organic cladding materials. In combination with simple camera read-out schemes, our concept shows a route towards disposable sensor chips for highly multiplexed detection of a wide range of analytes in point-of-care diagnostics.

## 2. $Si_3N_4$-organic hybrid device concept and design considerations

The basic concept of $Si_3N_4$-organic hybrid (SiNOH) lasers [20] is illustrated in Fig 1. The SiNOH lasers are optically pumped from top with an external light source, see Fig 1(a). A high-precision alignment of the pump spot is not required. Laser light is emitted directly into a $Si_3N_4$ single-mode WG and can be efficiently coupled to a subsequent sensor, here illustrated as a laser array feeding an array of Mach-Zehnder interferometers. At the sensor outputs, light is radiated into free space via grating couplers and captured by a camera. This concept allows contact-less excitation and the readout of large-scale sensor arrays while keeping the complexity of the chip low.

### 2.1 Device concept

The SiNOH WG have a $Si_3N_4$ core on top of a silicon dioxide bottom cladding. Optical gain is provided by an organic top cladding that is doped with a light-emitting dye. By an appropriate choice of the gain medium and of the device design, various emission wavelengths $\lambda_{1...n}$ can be realized on the same chip, potentially pumped with the same light source. Note that similar concepts for on-chip lasers emitting in the NIR have been previously demonstrated on the silicon platform [21]. In our experiments, we investigate two laser resonator types: A spiral-shaped ring resonator, Fig 1(b), and a distributed-feedback (DFB) structure, Fig 1(c). The ring resonator is coiled up into a long spiral and couples evanescently to a single-mode output WG, Fig 1(b). Compared to the DFB laser, the spiral covers a much larger chip area, and hence provides a large overlap with a moderately focused pump beam that leads to relaxed alignment tolerances. The distributed-feedback resonator is formed by a straight strip WG with sidewall corrugations, Fig 1(c). The footprint of the DFB laser resonator is small; therefore many SiNOH lasers can be pumped simultaneously with one pump beam. The DFB structure supports lasing in basically two longitudinal modes spectrally located at both sides of the Bragg wavelength. Asymmetries or a quarter-wavelength defect reduces the number of lasing modes to one. Pump efficiency, threshold and spectral properties depend strongly on the modal gain and on the loss of the respective laser types. This is discussed in the next subsections.

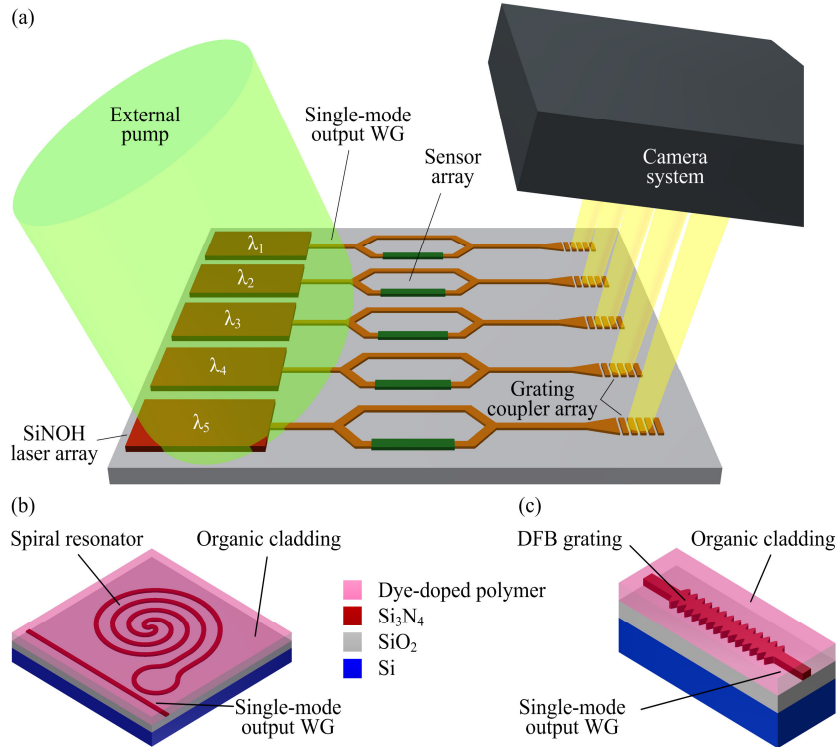

Fig. 1. Basic concept of integrated Si$_3$N$_4$-organic hybrid (SiNOH) lasers. (a) SiNOH lasers feeding an array of waveguide-based sensors. The lasers are optically pumped from top by an external light source with an extended spot, without any need for high-precision mechanical alignment. Laser light is emitted directly into the Si$_3$N$_4$ single-mode WG and can hence be efficiently coupled to an array of on-chip sensors, here illustrated as Mach-Zehnder interferometers. The light from the sensor output is coupled out via grating couplers and captured by a camera. (b) SiNOH laser with spiral-shaped resonator. The WG consists of a Si$_3$N$_4$ core with a dye-doped organic cladding as a gain medium. An appropriate resonator design in combination with a suitable gain medium allows to realize various emission wavelengths $\lambda_{1\ldots n}$ on the same chip. The ring resonator is coiled up into a spiral and couples evanescently to a single-mode output WG. (c) SiNOH laser with distributed-feedback resonator, formed by a strip WG with sidewall corrugations.

## 2.2 Gain and loss in SiNOH WG

The power of the oscillating laser mode depends on the SiNOH WG loss and on the efficiency with which the pump is coupled to the active WG region. Assuming effective propagating properties, the waves experience a net power gain of $G = \exp\left[(g-\alpha)2L\right]$ for one round-trip of length $2L$ in the resonator. In the case of a ring resonator, $2L$ denotes the ring perimeter, while for the case of a DFB laser $L$ represents the geometrical resonator length. The modal power gain and loss coefficients are denoted by $g$ and $\alpha$, respectively (unit $\mathrm{cm}^{-1}$). When lasing, the power gain is $G = 1$, i. e. $g = \alpha$.

The gain of the laser resonator results from the interaction of the evenanescent parts of the WG mode with the dye molecules in the organic top cladding, which are excited by the pump light. The simulated electric field magnitudes $|\vec{E}|$ of the quasi-TE and quasi-TM modes in the cross-section of a strip WG are shown in Fig. 2(a) and (b), respectively.

**Gain and loss in ring resonators**   Loss of the laser mode is caused by attenuation of the mode when propagating along the ring, and by the power output coupling $|\kappa_\mathrm{r}|^2$ to the bus WG [23]. The net roundtrip power gain and the net modal gain at threshold is therefore

$$G = \mathrm{e}^{(g-\alpha)2L}\left(1-|\kappa_\mathrm{r}|^2\right) = 1, \qquad (g-\alpha)2L = -\ln\left(1-|\kappa_\mathrm{r}|^2\right). \tag{1}$$

**Gain and loss in DFB resonators**   The resonator mode can be described by two counter-propagating waves. The grating couples these fields with a coupling factor $\kappa$. This coupling factor is real, because the penetration depth of the evanescent field follows the grating closely so that the amplifying volume remains virtually constant along the grating, Fig. 2(d),(e). For low gain, $\frac{1}{2}(g-\alpha) \ll \kappa$, the first order resonances of a DFB laser are at the stopband edges,

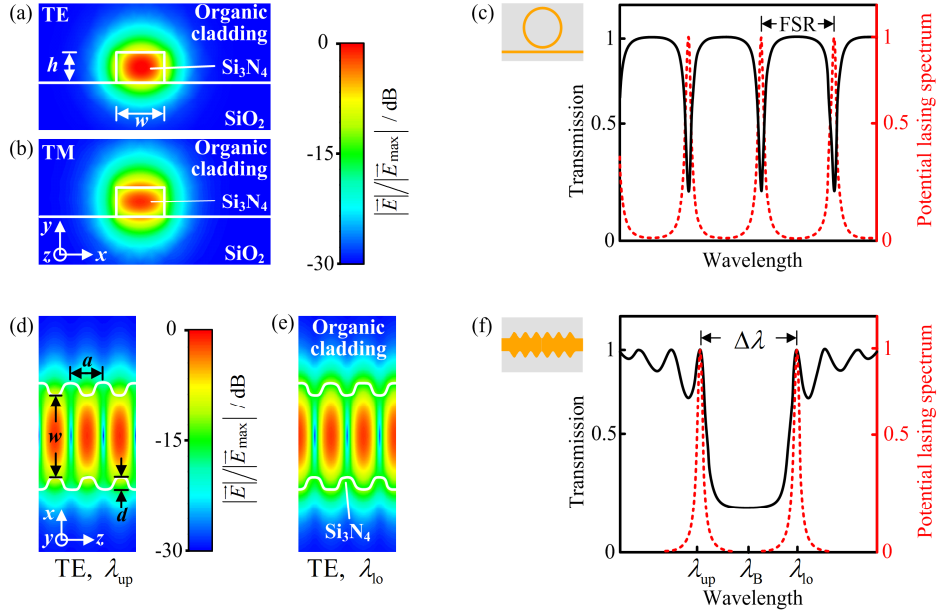

Fig. 2. Coiled-up Si$_3$N$_4$ ring resonator (spiral resonator) and Si$_3$N$_4$ DFB resonator on a SiO$_2$ substrate. The WG cores are clad with an organic material. Simulated electric field magnitudes $|\vec{E}|$ and spectral transmission of the laser resonator. Field propagation is along the $z$-axis.   (a), (b) Spiral resonator. Cross-section of strip WG with width $w$ and height $h$. Fundamental quasi-TE mode (a), and fundamental quasi-TM mode (b).   (d), (e) DFB resonator: Top view of corrugated strip WG with width $w$, height $h$, corrugation period $a$, and corrugation depth $d$. The fundamental quasi-TE mode fields are plotted for $y$ = const. at half the WG height. Upper band edge with larger photon energy and short-wavelength resonance $\lambda_\mathrm{up}$ (d), lower band edge with smaller photon energy and long-wavelength resonance $\lambda_\mathrm{lo}$ (e).   (c) Spiral resonator: Schematic power transmission between the input and the output of the straight bus WG (black solid line) as a function of the vacuum wavelength $\lambda$. Lasing occurs at the resonances (red dashed line) marked by notches in the transmission.   (f) DFB resonator: Schematic power transmission between the input and the output of the corrugated section (black solid line) as a function of the vacuum wavelength $\lambda$. Lasing occurs at the resonances closest to the Bragg wavelength $\lambda_\mathrm{B}$ (red dashed lines).

and the net modal threshold power gain can be approximated by [22, Eq. (29), (6)]

$$(g - \alpha)L = 2\left(\frac{\pi}{\kappa L}\right)^2 . \tag{2}$$

### 2.3 Design considerations

For lasers operated in sensor applications, the important characteristics are the spectrum, the polarization, the threshold and the output power. These parameters are defined by the resonator, the properties of the laser gain medium, and the pump.

**Spectrum and polarization**  The organic gain medium is dominantly inhomogeneously broadened. If there is no longitudinal mode filtering, the laser emission wavelengths are determined by the spectral dependence of the laser dye emission as depicted in Fig. 3(d), red curve.

With a ring-shaped laser resonator, the emission spectrum of the organic laser dye is longitudinally filtered according to the comb-shaped transmission spectrum with a free spectral wavelength range $FSR = \lambda_0^2/(n_{eg}L)$ centered at the operating wavelength $\lambda_0$, Fig. 2(c). In addition to the fundamental TE and TM mode, Fig. 2(a)(b), transverse modes of higher order can propagate in wider WG ($w \geq 400\,\text{nm}$). For the ring resonator we find one set of longitudinal modes for each transverse mode in each polarization. The amount of lasing modes can be reduced with additional filters.

For a DFB grating with period $a$, the refractive index modulation along the waveguide can be described by an effective index modulation $n_e(z) = n_{av} + \Delta n \sin(2\beta_B z)$, an amplitude $\Delta n$, and an average effective refractive index $n_{av}$. The propagation constant is $\beta_B = 2\pi n_{av}/\lambda_B$ at the Bragg wavelength $\lambda_B = 2n_{av}a$. The coupling strength of the grating for weak coupling is given by [22]

$$\kappa = \frac{\pi \Delta n}{\lambda_B} . \tag{3}$$

The transmission spectrum of a DFB structure and its associated stopband can be seen in Fig. 2(f), black line. The transmission bandgap is centered at the Bragg wavelength $\lambda_B = 2n_{av}a$. For a grating with length $L = ma$, $m$ is the amount of grating periods. The bandwidth $\Delta\lambda$ of the stopband for a first order Bragg grating is given by [22]

$$\Delta\lambda = \frac{\lambda_B^2}{\pi n_{av}}\sqrt{\kappa^2 + \frac{\pi^2}{L^2}} . \tag{4}$$

In a DFB grating, laser emission does not occur at $\lambda_B$ in the center of the stopband. However, two first-order longitudinal resonant modes are located at the stopband edges, Fig. 2(f), red dashed line, and lead to laser oscillation. If a quarter-wavelength shift is included in the center of the Bragg grating, i. e., if a defect is introduced in the periodic structure, laser emission occurs only at the Bragg wavelength, where the resonator loss is minimum.

As with the ring resonator, we find one set of longitudinal modes for each transverse mode in each polarization.

**Laser threshold and output power**  The estimation for the net modal threshold power gain, Eq. (1),(2), holds for laser modes in inhomogeneously broadened gain media, as longitudinal modes do not compete for gain. The organic gain medium emission is inhomogeneously broadened, while in a sufficiently small wavelength range the transitions are homogeneously broadened. Within that range all longitudinal modes compete for gain and finally the mode with the largest net gain oscillates. The more modes are competing, the higher the threshold for the dominant mode becomes. The two lasing modes of DFB lasers without quarter-wavelength shift are separated sufficiently so that no competition occurs. The modes of ring resonators with large diameters have a small FSR, and therefore many closely neighbored longitudinal modes exist with consequently high thresholds. An additional longitudinal mode filter reduces the number of longitudinal modes inside the homogeneous linewidth and reduces the pump threshold.

TE and TM polarized modes have a large spatial overlap inside the gain medium, Fig. 2(a),(b). Therefore laser modes with similar frequencies but different polarizations compete for gain so that the threshold increases in these cases.

A high differential quantum efficiency $\eta_d$ of the laser is important for a high output power. For higher pump powers the organic laser dyes are degrading faster, and therefore the lifetime of the SiNOH laser is reduced. The ratio of induced and total emission (meaning induced and spontaneous emission) is denoted as $\eta_{ind}$. The spontaneous emission depends on the total number of modes inside the linewidth of the spontaneous emission and therefore on the volume of the pumped region. The total photon lifetime $\tau_P$ in the resonator, $\tau_P^{-1} = \tau_R^{-1} + \tau_\alpha^{-1}$, including resonator losses $\tau_\alpha^{-1} \propto \alpha$ and outcoupling loss $\tau_R^{-1}$, is to be compared to the lifetime $\tau_R$ due to outcoupling only. The differential quantum efficiency is then given by

$$\eta_d = \eta_{ind} \frac{1/\tau_R}{1/\tau_P}. \qquad (5)$$

Low propagation loss $\alpha$ requires a low threshold modal gain $g$, Eq. (1), (2), and leads to a high quantum efficiency for a given outcoupling, Eq. (5). The outcoupling has to be optimized for extracting the maximum power from the resonator. A small active volume with a small number of modes which can accept spontaneous emissions results in a larger $\eta_{ind}$ than a larger active volume, if the gain in both cases is equal.

DFB lasers have a small footprint, and therefore the active region has a small overlap with the pump spot so that the threshold is relatively large. According to Eq. (2), the threshold decreases with longer WG. A large overlap of the pump spot with a long spiral WG offers an efficient pumping of the organic gain material and leads to a low threshold gain, Eq. (1). The internal laser power is uniformly distributed along the spiral WG. However, as said before, the outcoupling must be optimized for the maximum output power in each case.

### 3. Fabrication

The SiNOH laser waveguides are fabricated with only one lithographical step. The WG cores are structured in a 200 nm thick $Si_3N_4$ layer, on top of a 2 μm thick silicon dioxide layer which is mechanically supported by a silicon wafer. The etch mask is a negative resist structured by electron beam lithography and spray developing. Dry etching with a mixture of $SF_6$ and $CHF_3$ is used to transfer the structure of the mask to the $Si_3N_4$ layer. An oxygen plasma etching step follows and removes the etch mask. After structuring the WG cores a 800 nm thick PMMA layer in which the laser dye PM597 is dispersed was spin coated and acts as the active cladding ($25 \mu M/g$, Pyrromethene 597, Radiant Dyes Laser & Accessories GmbH). For characterization, the chips are cleaved on one side thereby providing access to the facet of the bus WG.

Figure 3(a) shows a scanning electron microscopy (SEM) image of the structured $Si_3N_4$ WG cores of a spiral laser with ring resonator mode filter. The spiral WG is curled-up densely with a minimal curvature radius of 40 μm for low radiation losses. Figure 3(b) displays a part of a strip WG with sidewall corrugations. Figure 3(c) shows a cross-section of the $Si_3N_4$ WG core embedded in the dye-loaded PMMA cladding.

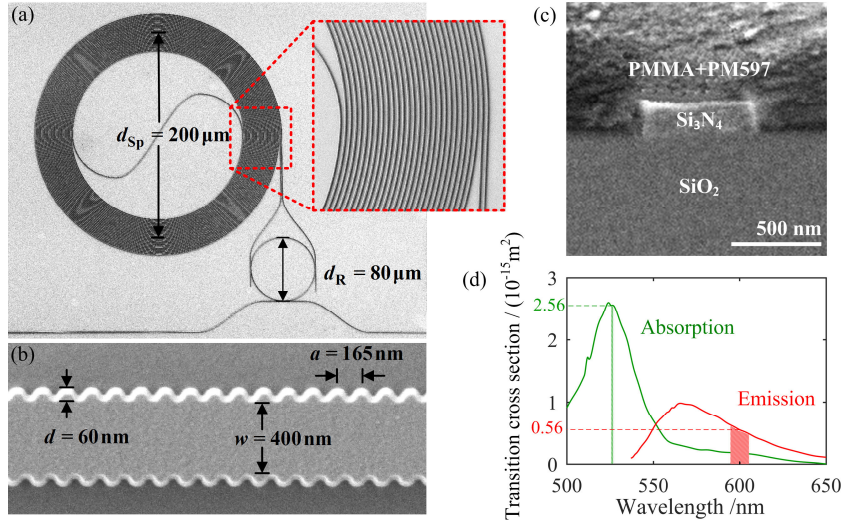

Fig. 3. SEM images of the $Si_3N_4$ WG with absorption and emission spectra of the organic cladding material. (a) Spiral resonator: The spiral WG is densely coiled-up with a minimal diameter of 80 μm for keeping the radiation losses small. (b) DFB resonator: Corrugated WG with width $w$, corrugation depth $d$, and period $a$. (c) Cross-section of the $Si_3N_4$ WG core clad with PMMA and Pyrromethene (PM597). (d) Emission and absorption cross-sections of PM597 embedded in PMMA. The material is laser-pumped near the absorption maximum of PM597 in PMMA, green line ($\sigma_a$(532 nm) = 2.56·$10^{-15}$ $cm^2$). The emission of the SiNOH laser is near 600 nm, red hatched area ($\sigma_e$(600 nm) = 0.56·$10^{-15}$ $cm^2$). It is redshifted from the emission maximum, because of the stronger reabsorption for shorter wavelengths.

Figure 3(d) finally shows the absorption and emission cross sections of PM597 dispersed in PMMA. The SiNOH laser is pumped near the absorption maximum, Fig. 3(d), green line. The emission of the SiNOH laser appears at around 600nm wavelength, redshifted from the maximum of the emission of the laser dye, red line. This can be explained by the higher probabilities for re-absorption, marked by an overlap of the absorption and emission spectra. Moreover, for shorter wavelengths also higher-order transverse modes can be guided, leading to mode competition and a larger threshold gain for each single laser mode.

## 4. Characterization and demonstration of lasing

We first investigate the influence of the resonator geometry on the laser threshold and on the laser spectrum in a setup according to Fig. 4. In subsequent sections we then discuss the operating properties of the spiral and of the DFB laser.

### 4.1 Characterization setup

The SiNOH lasers are pumped from top by a frequency doubled Nd:YLF pulsed laser with 523 nm wavelength, a pulse duration of 20 ns and a repetition rate of 20 Hz (CL523, CrystaLaser), Fig. 4. The pulse energy is varied with a half-wave ($\lambda/2$) plate and a polarizer $P_1$, and a beam splitter BS transmits half of the pump energy to an energy meter. The pump light is focused on the chip by a lens $L_1$. The SiNOH laser light is coupled from the chip edge with a polarization maintaining (PM) lensed fiber. A collimator $C$ and a refocusing lens $L_2$ direct the emitted light to a spectrometer (Shamrock 500i, Andor, 60 pm resolution, 2 s integration time). In the case of a spiral laser, an optional rotatable linear polarizer $P_2$ is inserted between the fiber collimator and the refocusing lens for selecting the proper polarization of the lasing mode. An auxiliary alignment laser and a grating coupler help in actively adjusting the lensed fiber to the output WG of the SiNOH laser. The transmission of the DFB grating is measured by coupling the light of a white light source (SuperK, NKT Photonics) via the grating coupler, and by capturing the transmitted light at the output with the lensed fiber which illuminates the spectrometer.

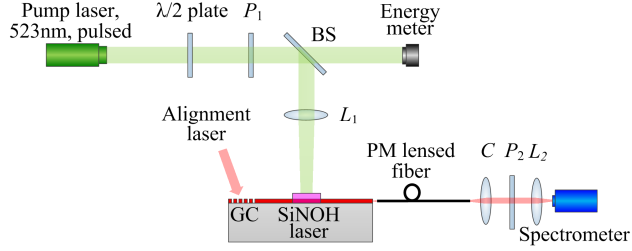

Fig. 4. Measuring the threshold and the spectrum of SiNOH lasers. The SiNOH laser is pumped with a free-space beam from a pulsed laser which is focused by a lens $L_1$ on the SiNOH laser. The pulse energy of the pump laser is varied by rotating a half-wave plate with respect to a fixed linear polarizer $P_1$. A beam splitter BS transmits half of the pump emission to an energy meter. The SiNOH laser emission is coupled from the chip by a lensed polarization-maintaining (PM) fiber. A collimator $C$ and a refocusing lens $L_2$ direct the light to a spectrometer. When measuring a spiral laser, a rotatable linear polarizer $P_2$ is inserted in the collimated portion of the beam for selecting the proper polarization of the lasing modes. The lensed fiber is adjusted with the help of an auxiliary alignment laser at 635 nm which is coupled to the bus WG (spiral laser) or to the corrugated WG (DFB laser) through a grating coupler.

### 4.2 Spiral laser

The ring resonator has a perimeter of $L = 17.5\,\text{mm}$ and consists of a densely packed double spiral for optimally filling the pump spot area, Fig. 5(a)-(c). The pump beam has a Gaussian profile with a full-width half-maximum spot size of 200 μm. The curvature of the spiral-shaped WG is limited to a minimum radius of 80 μm by the bending loss which can be tolerated. To avoid high propagation loss for the fundamental TE and TM modes, the WG are multimoded with a width of $w_\text{MM} = 500\,\text{nm}$ (MM, black lines in Fig. 5(a)-(c)), leading to a loss of 4.6 dB/cm for the straight WG sections. This WG width implies a stronger confinement of the optical field to the WG and therefore a lesser overlap with the gain medium. Thus one would expect a smaller gain than could be achieved with narrower WG. However, we found that the single-moded WG have a larger loss of 6.7 dB/cm, leading to 10 to 20 times larger thresholds than for multimoded spirals. We suppress the oscillation of higher order transverse modes by single-mode filters (orange lines in Fig. 5(a)-(c)), which are inserted in the center of the spiral (curvature radius 40 μm) and as a loop in the coupling section between spiral and straight bus WG.

In a first step, we measure the lasing threshold of the spiral laser with a loop at the end of the spiral WG, Fig. 5(a). The (green) pump spot overlaps partially with the spiral's evanescent field which reaches into the laser-active material. However, much of the pump energy remains unused. Figure 5(d) shows the normalized lasing spectrum, integrated over $2\,\text{s}$ and for a pump pulse energy of $715\,\text{nJ}$. The FSR of the spiral-shaped ring is $\text{FSR}_\text{Spiral} = 10\,\text{pm}$ corresponding to 8.3 GHz (spectrometer resolution 60 pm). Lasing occurs only in a small range ($590\,\text{nm}$ to $610\,\text{nm}$) of the spontaneous emission spectrum of the laser dye, but the center of the lasing spectrum changes from integration interval to integration interval. Possibly mode competition could play a role, but the true reason is not yet known. We further integrate over all spectral line shapes and depict the calculated normalized output power as a function of the pump pulse energy $W$, Fig. 5(g). No significant fluctuation of the total power is observed which supports the idea of mode competition. We find a laser threshold of 312 nJ, corresponding to a pump fluence of $400\,\mu\text{J}/\text{cm}^2$ for the area occupied by the spiral WG.

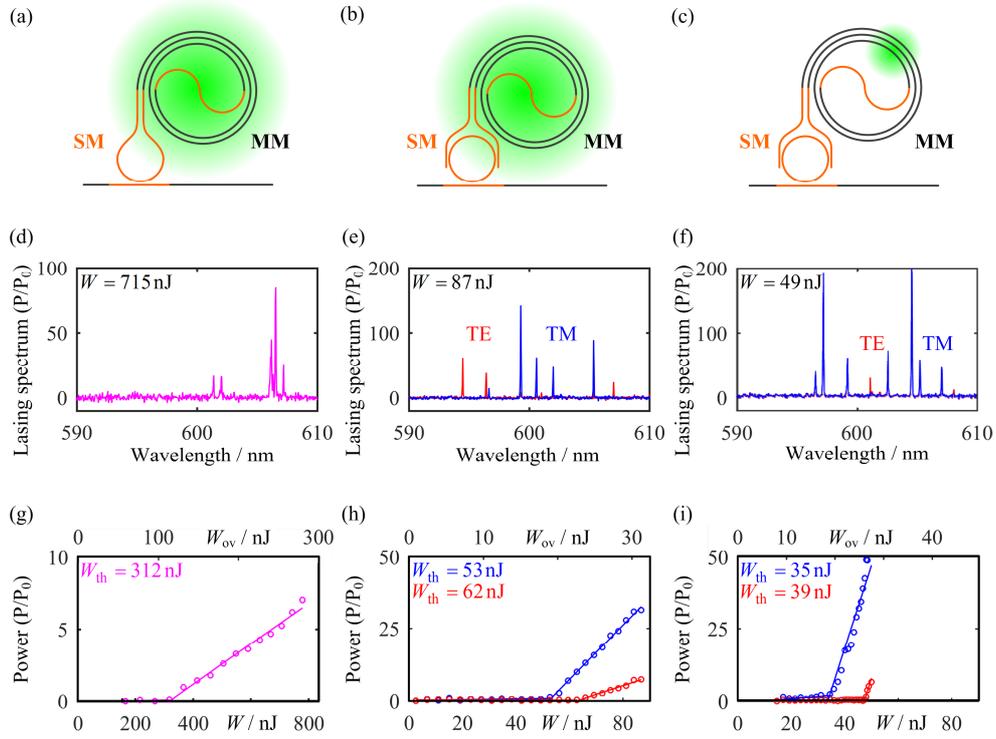

Fig. 5. Spiral laser. Threshold and lasing spectra for different mode filters and pump spot sizes. (a)-(c) Resonator geometries and pump spots (green). A transverse mode filter (orange) is implemented by tapering down the multimode spiral-shaped resonator WG from 500 nm width to 300 nm for single-mode propagation. (b),(c) A coupled ring resonator acts as a longitudinal mode filter and reduces the number of lasing modes in the spiral resonator. A pump spot size in the order of the spiral diameter wastes pump energy (a),(b). Focusing the pump spot to an area with high WG density (c) decreases the threshold. (d)-(f) Emission spectra. In (d) the scaling of the vertical axis differs from the scaling in (e),(f). If the small ring resonator is present, its FSR determines the lasing lines. (g)-(i) Laser output power as a function of the total pump pulse energy $W$ (theshold energy $W_{th}$). The pump has a Gaussian spot which (partially) overlaps with the area occupied by the spiral resulting in an overlap pump energy $W_{ov}$, upper horizontal axis. In (g) the scaling of the axes differs from the scaling in (h),(i). An additional ring filter with a round trip propagation loss of 1 dB (4 dB/cm, $Q = 5\times10^3$) decreases the theshold by a factor of 6 compared to (g), where only the spiral with a round trip propagation loss of 6 dB exist (3.5 dB/cm). By enhancing the overlap of the pumpspot with the spiral the threshold further decreases by a factor of 1.5, and the slope of the laser characteristic is increasing.

Next we investigate a spiral laser, where the coupling loop of Fig. 5(a) is expanded to be a small ring resonator with $FSR_{Ring} = 660\,\mathrm{pm}$ corresponding to 540 GHz, Fig. 5(b). The pump overlap with the laser-active volume is the same as in Fig. 5(a). Figure 5(e) shows the lasing spectrum for a pump pulse energy of 87 nJ. Lasing occurs for the fundamental TM mode (blue), and for the fundamental TE mode (red) according to multiples of $FSR_{Ring}$. The spectrum differs greatly from Fig. 5(d) and it is stable from integration interval to integration interval. We attribute this to the fact that the homogeneous line width of the dye is smaller than $FSR_{Ring}$ so that different modes to not compete for gain. As a consequence, the pump energy at threshold, Fig, 5(h), is smaller than for the case of modes competing for gain as in Fig. 5(a), (d), (g). Figure 5(h) shows further that TM-polarized lasing modes have lower thresholds (53 nJ) than TE-polarized modes (62 nJ), corresponding to pump fluences of $56\,\mathrm{\mu J/cm^2}$ (TM) and $65\,\mathrm{\mu J/cm^2}$ (TE) at the laser-active area of the spiral. This is attributed to the fact that TM-polarized modes transport a larger fraction (0.2) of cross-sectional power in the gain medium outside the WG core than the TE modes (0.17). In addition, the loss for a TM-polarized mode (high electric field strengths at the smooth upper surface) is smaller than for a TE-polarized mode (high electric field strengths at the rough sidewalls).

The differential quantum efficiency Eq. (5) increases significantly as is seen by comparing the slopes in Fig. 5(g) and Fig. 5(h). The reason is that the photon lifetime for the combination of filter ring and spiral in Fig. 5(h) is much larger than for the spiral of Fig. 5(g) alone, while the outcoupling to the bus WG is the same in both cases. The ratio $\eta_{ind}$ of induced and total emission in Eq. (5) sees a comparable mode number in both cases

The threshold of the spiral laser can be further decreased by increasing the overlap of the pump spot with the spiral WG. One way is to focus the pump spot to an area with densely spaced WG, Fig. 5(c), green spot, so that the effective pump fluence increases. The emission spectrum shows similar properties compared to Fig. 5(e). The lasing threshold, however, is 1.5 times smaller for the focused pump spot, and the corresponding pump fluence at threshold is $980\,\mu J/cm^2$, Fig. 5(i). However, this is not a fair comparison, because the pump energy is more efficiently used in the setting of Fig. 5(c) than in the one of Fig. 5(b).

We therefore better discuss the fraction $W_{ov}$ of the pump pulse energy that overlaps with the spiral WG, upper horizontal of Fig. 5(h),(i). With respect to $W_{ov}$ the threshold for both cases is the same, and this confirms the conclusion from Fig. 3d that the reabsorption in the unpumped regions is small. However, the differential quantum efficiency is larger for the focused pump beam, because the active volume is smaller and $\eta_{ind}$ therefore larger, see text after Eq. (4).

### 4.3 Distributed feedback laser

Sidewall corrugations of a waveguide scatter light and establish a distributed feedback (DFB) resonator, Fig 6(a). The DFB structure has a length $L$, a width $w$, and the corrugations have a depth $d$ and a period $a$. The feedback strength is $\kappa$, Eq. (3), and the separation of the lowest-order lasing modes is $\Delta\lambda$, Eq. (4). The active medium of the DFB laser is pumped from top with an elliptical pump spot.

Figure 6(b), black line, shows the typical transmission spectrum of a passive DFB resonator measured by coupling the light of a super continuum laser (SuperK Versa, NKT Photonics GmbH) to the GC, and detecting the transmitted light with the spectrometer, Fig. 4. The resolution bandwidth (RBW) is as large as 260 pm, because the super continuum source is limited in power. As a consequence, the transmission maxima at the band edges are blurred. The emission spectrum of the pumped resonator is depicted in the same figure, red line and shows two lasing modes with $\Delta\lambda = 1.2\,nm$ spectral distance. The induced gain and a temperature rise leads to a red shift as compared to the cold resonator. For better visualization the transmission spectrum was redshifted by $0.2\,nm$. A dominant mode occurs at the lower bandgap edge for lower photon energies, $TE_{lo}$, whereas the upper bandgap lasing mode, $TE_{up}$, is less pronounced.

The laser output power for and was measured as a function of the pump pulse energy W, Fig. 6(c). The dominant laser mode of Fig. 6(b), filled red circles, starts lasing for lower pump pulse energies $W_{th,lo} = 32\,nJ$ than the less pronounced one with $W_{th,up} = 41\,nJ$, open red circles. The differential quantum efficiency Eq. (5) is larger for the $TE_{lo}$ mode than for the $TE_{up}$ mode. In an ideal DFB laser both modes would have the same net gain und therefore the same threshold and the same differential quantum efficiency. Due to fabrication inaccuracies the corrugation is positioned slightly asymmetrically with respect to any end face reflections so that one of the modes becomes dominant. As both modes are amplified, the homogeneous line broadening of the dye must be smaller than the spectral separation of the DFB resonator modes.

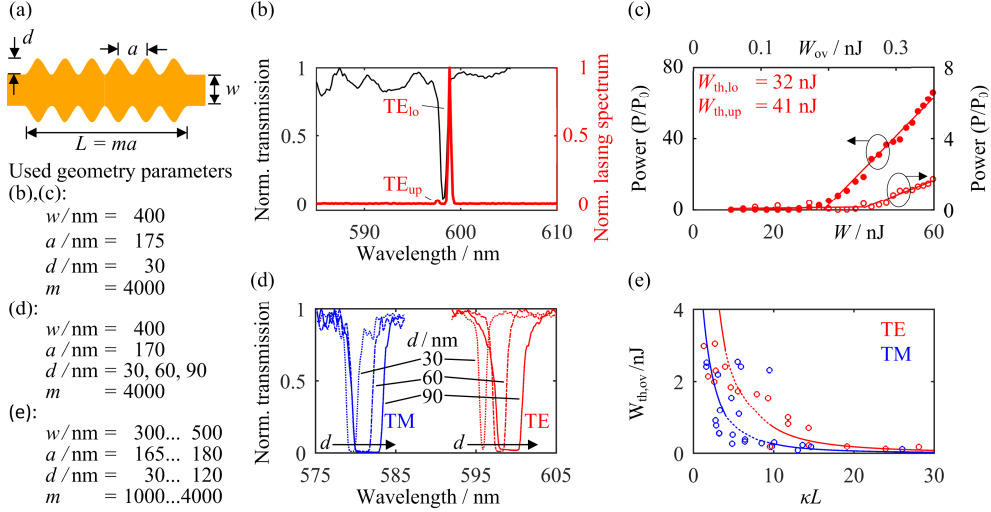

Fig. 6. Transmission and threshold for DFB resonators with various geometries. (a) Schematic of a DFB section with length $L$, corrugation depth $d$, period $a$, and width $w$. Parameters for Fig. 6(b)-(e). (b) Normalized DFB WG transmission (black line, RWB = 260 pm) and normalized spectrum (red line, RBW = 60 pm) of a DFB laser. The lasing modes are located at the edges of the stopband which are blurred because of the coarse resolution. The dominant laser mode occurs at the lower (in frequency terms) stopband edge $TE_{lo}$. The lasing mode at the upper stopband edge $TE_{up}$, has much less power. (c) Laser characteristics. The dominant laser mode $TE_{lo}$ starts lasing for lower pump pulse energies (●, $W_{th,lo}$ = 32 nJ, left vertical axis) than the weak mode $TE_{up}$ at the upper stopband edge (○, $W_{th,up}$ = 41 nJ, right axis). The horizontal upper axis gives the fraction of the pump pulse energy $W_{ov}$ that overlaps with the DFB structure. (d) Normalized DFB WG transmission for TE (red lines) and TM polarization (blue lines) for different corrugation depths $d$. The larger $d$ becomes, the wider the stopband width $\Delta\lambda$ is. $\Delta\lambda$ is larger for TM compared to TE polarization for the same corrugation depth $d$. (e) Threshold pump pulse energies for different coupling strengths $\kappa L$ for TE and TM polarization. Only the fraction of the energy $W_{ov}$ of the pump beam that overlaps with the geometrical DFB structure is considered. The laser threshold decreases for larger $\kappa L$. For TM polarization it decreases faster than for TE polarization. This can be explained by the larger overlap of the TM mode with the gain medium and by the lesser propagation loss compared to the TE polarized mode.

Figure 6(d) shows transmission spectra for DFB resonators with different corrugation depths $d$, having otherwise the same geometry as in Fig. 6(b) for TE (red lines) and TM modes (blue lines). Thus the coupling strength $\kappa$ was varied, but also the average WG width $w_{av} = w + d/2$. The bandgap centers for TE modes occur (16…18) nm redshifted from the ones of TM polarized modes. The width $w$ of the WG core is 2 times larger than its height. This leads to a larger confinement and therefore to a larger $n_{av}$ for TE polarized modes. With increasing $d$ the average WG width $w_{av}$ and the effective index $n_{av}$ increase. This increase is stronger for TE than for TM modes. As a consequence, the center wavelength $\lambda_B = 2n_{av}a$ increases stronger for TE than for TM modes. The bandwidth $\Delta\lambda$ is slightly larger for TM polarized modes than for TE polarized modes for the same $d$. This means that the grating strength is larger for TM modes.

The threshold pump pulse energies for different coupling strengths $\kappa$ and different lengths $L=ma$ are shown in Fig. 6(e) for TM polarization, blue circles, and for TE polarization, red circles. From simulated effective indices for a homogeneous WG with widths $w$ and $w + d$ we calculate $\Delta n$ and the coupling strengths $\kappa$ according to Eq. (3). For a better comparison, only the fraction of the pump pulse energy $W_{th,ov}$ is considered that overlaps with the DFB resonator. With the assumption $W_{th,ov} \propto (g-\alpha)L$, guidelines were vertically fitted for both polarizations according to Eq. (2) for low gain ($\kappa L$ large). With the same vertical scaling we used the numerical solution [22, Fig. 8] for high gain ($\kappa L$ small). A dashed spline connects both partial curves.

The laser threshold decreases for larger $\kappa L$. For TM polarization it decreases faster than for TE polarization. This can be explained by the larger overlap of the TM polarized mode with the gain medium compared to the TE polarized mode. In addition, the propagation loss is expected to be smaller for TM polarized modes as large field strengths are located at the smooth upper surface of the WG core. For TE polarized modes, large field strengths are located at the rather rough sidewalls of the WG core. Low thresholds of $W_{\text{th,ov}} < 0.3\,\text{nJ}$ are reached with total coupling strengths $\kappa L > 15$. The pump fluences on the resonator at threshold are $(40...60)\,\mu\text{J}/\text{cm}^2$.

Note that a lower threshold does not necessarily lead to higher differential quantum efficiency. If the feedback is very strong, $\tau_R$ becomes much larger than $\tau_\alpha$, Eq. (5), i. e., all light will be confined in the laser resonator, and there is no efficient outcoupling. Nevertheless, this effect could not be seen within our resonator study.

In order to achieve single mode lasing, DFB lasers with a quarter-wave shift ($\text{DFB}_{\lambda/4}$) at $L/2$ were investigated. The half-period ($\lambda/4$) shift in the center of the grating induces a $\pi/2$ phase shift, which leads to a lasing at the Bragg wavelength $\lambda_B$. Figure 7(a) shows a schematic of a DFB section with quarter-wave shift in the middle, with total length $L$, corrugation depth $d$, period $a$ and width $w$.

Figure 7(b), black line (RBW = 260 pm), shows the typical transmission spectrum of a passive $\text{DFB}_{\lambda/4}$ resonator with $\kappa L = 3.4$. The emission spectrum of the pumped resonator is depicted in the same figure, red line, RBW = 60 pm. Although there is no characteristic transmission peak at $\lambda_B$ in the middle of the transmission stopband, the laser mode occurs at $\lambda_B$. Because for the white-light measurement the resolution of the spectrometer could not be chosen sufficiently high, the transmission peak in the stopband cannot be seen in the black curve.

The laser output power was measured as a function of the pump pulse energy $W$, Fig. 7(c). The single laser mode starts lasing for pump pulse energies $W_{\text{th}} = 33\,\text{nJ}$ ($W_{\text{th,ov}} = 0.07\,\text{nJ}$), which is comparable to the threshold of the DFB gratings with large $\kappa L$ in Fig. 6(d). The pump fluence on the resonator at threshold is $64\,\mu\text{J}/\text{cm}^2$.

Besides the lasing mode located at $\lambda_B$, the two "DFB lasing modes" located at the edges of the bandgap, Fig 6(b), can be amplified with lesser gain, if there is no competition for gain. As both "DFB modes" are not amplified for higher pump energies, the homogeneous line broadening of the dye must be larger than the spectral separation of the $\text{DFB}_{\lambda/4}$ modes. From that and our findings from the DFB lasers, we can set a lower limit $\Delta\lambda_{lo} = 0.5\,\text{nm}$ and an upper limit for the homogeneous line broadening of the gain medium, $\Delta\lambda_{up} = 1.2\,\text{nm}$.

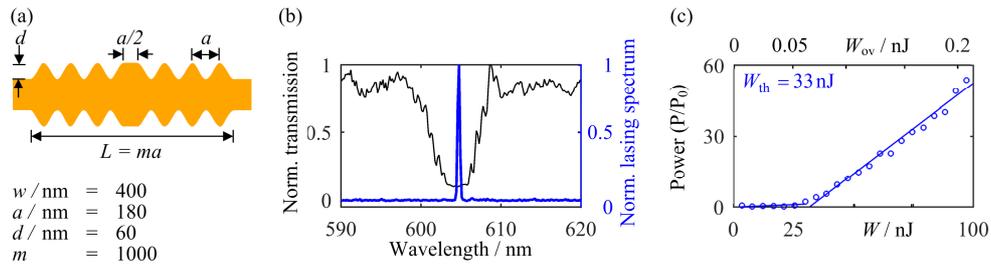

Fig. 7. Transmission spectrum and threshold of a $\text{DFB}_{\lambda/4}$ laser. (a) Schematic of a DFB section with λ/4 shift at $L/2$, total length $L$, corrugation depth $d$, period $a$, and width $w$. The half-period (λ/4) shift in the center of the grating induces a π/2 phase shift, which leads to lasing at the Bragg frequency. Parameters for Fig. 6(b),(c). (b) Normalized transmission spectrum (black line, RBW = 210 pm) and laser emission (blue line, RBW = 60 pm) for a TM polarized mode. The lasing mode is located at $\lambda_B$ in the middle of the stopband. The expected transmission peak at $\lambda_B$ cannot be see, because the resolution of the spectrometer could not be chosen sufficiently large. (c) Laser characteristics: The TM laser mode starts lasing for pump pulse energies $W_{\text{th}} = 33$ nJ. The horizontal upper axis shows the fraction of the pump pulse energies $W_{\text{ov}}$ that overlaps with the $\text{DFB}_{\lambda/4}$ laser.

## 5. Discussion

After analyzing two different basic laser geometries we compare the properties that are crucial for biophotonic applications. Important requirements are the emission bandwidth, the polarization, the total emitted power, the resilience with respect to external feedback, and the compactness on the chip. The laser geometries and their properties are summarized in Tab. 1 and discussed in the following section.

**Tab. 1. Comparison of laser properties for different geometries.**

|  | # lasing modes / total bandwidth | Polarization fundamental mode | Resilience with respect to external feedback | Total emitted power | Compactness |
|---|---|---|---|---|---|
| Spiral | 3…8 / ~ 15 nm | Mixed TE and TM | ++ | ++ | – |
| DFB | 1 or 2 / ~ 2 nm | Either TE or TM | + | + | ++ |
| $\text{DFB}_{\lambda/4}$ | 1 / $\leq 60\,\text{pm}$ | Either TE or TM | – | + | ++ |

**Number of modes and polarization** For the spiral resonator the number of modes is mainly defined by the FSR of the filtering and the spectral distribution of the net gain. In our experiments we found that lasing occurs at 600 nm within a 15 nm bandwidth. The number of modes within the bandwidth can be controlled by the radius of the filter ring. Note that smaller ring radii could lead to higher propagation losses and therefore to higher thresholds. The possible increase of the number of simultaneously oscillating modes (with larger ring radii) is limited by the minimum mode separation set by the homogeneous linewidth of the gain medium. The polarization is mixed TE and TM and could be distributed to two separate WG on the chip.

For the DFB resonator the emission wavelength can be chosen in a large spectral range, limited by the net gain spectrum. By carefully choosing the period of the grating the polarization can be controlled. Two modes can be amplified, but usually one mode is dominant. The modes are spectrally separated by ~ 2 nm. This bandwidth can be slightly controlled by varying the total coupling strength of the grating.

For $\text{DFB}_{\lambda/4}$ lasers, only one mode at the Bragg wavelength oscillates. Similar to the DFB lasers, the mode can be chosen in a large spectral range by varying the period of the grating.

**Emitted power and Footprint** The larger the overlap of the laser WG with the pump spot is, the more energy can be transferred to the SiNOH laser. For spiral lasers this overlap is large and can be further increased by focusing the pump light to an area with densely spaced WG, but this comes at the cost of a large footprint on the chip.

It is possible to adapt the radius of the laser ring so that its FSR is slightly larger than the homogeneous broadened line of the gain medium. In this case all pumped dyes can be used for stimulated emission which leads to a high emitted power.

For the DFB and the $\text{DFB}_{\lambda/4}$ laser, the overlap with the pump spot is much smaller than for the spiral. An increase of the overlap is only possible with a strong alignment and a beam shaping strategy. However, the small footprint can be advantageous, if an array of DFB or $\text{DFB}_{\lambda/4}$ lasers is densely packed into a small area and pumped in parallel with the same pump spot.

The separation of the two DFB lasing modes is usually larger than the homogeneous broadened linewidth of the gain medium. Only dyes that emit in two homogeneously broadened lines that spectrally overlap with the DFB lasing modes contribute to the total stimulated emission. For the $\text{DFB}_{\lambda/4}$ laser only dyes emitting in one homogeneous broadened linewidth can contribute to stimulated emission. In both cases the total emitted power is smaller than for spiral lasers.

**Resilience with respect to external feedback**  Back reflection into laser resonators can disturb and even suppress the internal laser oscillation. As the SiNOH laser emits directly into the single mode outcoupling WG, back reflections can be easily coupled back to the SiNOH laser.

In our experiments we found that for the spiral lasers subsequent devices like grating couplers or splitters did not reduce the total emitted power. The power output of DFB lasers, however, became unstable if a subsequent optical circuit was not properly designed or showed fabrication imperfections. Especially the comparably short $\text{DFB}_{\lambda/4}$ lasers were often dominated by spuriously oscillating Fabry-Perot modes generated by the region between DFB grating and end facet of the coupling WG.

## 6. Summary

We demonstrate and investigate a new class of hybrid organic dye lasers emitting at 600 nm wavelength that can be monolithically integrated on the silicon nitride platform.

The SiNOH lasers are operated by optically pumping from top requiring only low alignment precision. The lasers can be fabricated in one lithography step by structuring the silicon nitride waveguides in parallel with other devices on the same chip, and by subsequently dispensing or spin coating the organic gain medium on top of the WG cores. The monolithic fabrication and the easy-to-handle operation principle make the SiNOH laser highly attractive for biophotonic applications.

We investigate two different laser geometries, spiral-shaped ring resonators and DFB resonators. Different approaches for improving the performance of the SiNOH lasers were investigated by modifying the resonator geometry and the pump strategy. All investigated SiNOH lasers offer low threshold fluences of $(40\ldots70)\ \mu\text{J}/\text{cm}^2$ .

The SiNOH lasers fulfil crucial demands for biophotonic applications. Spiral lasers are very robust against external feedback and have high total emitted powers. DFB lasers offer a small footprint and a small bandwidth that can be chosen in a large wavelength range.

## Funding

Alfried Krupp von Bohlen und Halbach Foundation, European Research Council (ERC Starting Grant 'EnTeraPIC', number 280145), Karlsruhe School of Optics and Photonics (KSOP).

## Acknowledgments

We acknowledge support by the Karlsruhe Nano Micro Facility (KNMF), a Helmholtz Research Infrastructure at Karlsruhe Institute of Technology.